\documentclass[]{spie}  

\usepackage{amsmath,amsfonts,amssymb}
\usepackage{graphicx}
\usepackage[colorlinks=true, allcolors=blue]{hyperref}
\usepackage{braket}

\title{An Interferometrically Robust Opto-Mechanical Coupler to Beam Polarisation}

\author[1,2]{Hayat Abbas}
\author[1,*]{Thomas Fernholz}
\affil[1]{School of Physics \& Astronomy, The University of Nottingham, University Park, Nottingham NG7 2RD, UK}
\affil[2]{Physics Department, Jazan University, Al Maarefah Rd, Jazan, Saudi Arabia}
\authorinfo{E-mail: thomas.fernholz@nottingham.ac.uk\\  www.coldatomsgroupnottingham.com}

\usepackage[numbers]{natbib}
\date{\today}
 
\begin{document} 
\maketitle

\begin{abstract}
In this work, we investigate a tool for hybrid quantum systems that implements a transducer to map small position changes of a micro-mechanical membrane onto the polarization of a laser beam. This is achieved with an interferometric setup that has reduced needs for stabilization. Specifically, an oscillating silicon nitride membrane placed in the middle of an asymmetric optical cavity causes phase shifts in the reflected, near-resonant light field. A beam displacer is used to combine the signal beam with a mode-matched, orthogonally polarized reference beam for polarization encoding. Subsequent balanced homodyne measurement is used to detect thermal membrane noise. Minor improvements in the design should achieve sufficiently high signal-to-noise ratio for the detection of motional quantum noise in the regime of high opto-mechanical coupling strength. 
This setup can provide a robust quantum link between a micro-mechanical oscillator and other systems such as atomic ensembles.
\end{abstract}

\maketitle

\section{Introduction}
Hybrid quantum systems have received significant interest, especially with the goal of technological exploitation of complementary capabilities for quantum information processing and communication tasks. Quantum transducers can be used to couple the properties of one object or system to different properties of another system, thus combining e.g.\ robust transmission of photonic quantum states with strong interactions between material quantum objects \cite{Kurizki2015}. Driven further, this type of research develops a toolbox for the engineering of strong interactions between quantum systems, thus building quantum machines.

A prototypical system for opto-mechanical coupling is the membrane-in-the-middle (MIM) arrangement \cite{thompson2008strong} where a micro-mechanical membrane couples to the electromagnetic field inside an optical resonator. 
A typical membrane material is silicon nitride ($\text{Si}\text{N}$), which combines low optical absorption in the near infrared with low mechanical loss \cite{wilson2009cavity}. Engineering of $\text{Si}\text{N}$ membranes and beams led to demonstrations of extremely high mechanical quality (Q) factors 
\cite{ norte2016mechanical,reinhardt2016ultralow,tsaturyan2017ultracoherent,ghadimi2018elastic} that enable
quantum mechanical experiments with massive objects at room temperature. There has also been a plethora of experiments on the quantum mechanical interaction between light and atomic ensembles \cite{Hammerer2010} where polarised light was used to drive effective interactions \cite {Muschik2013}, generate entanglement \cite{Krauter2011}, and deterministically teleport quantum states between macroscopic objects \cite{fernholz2013}. 
It is thus the combination of these systems that currently receives interest for the realisation of hybrid systems. It served for
studies of quantum measurement backaction in the optical detection of macroscopic objects and demonstration of backaction evading measurement of mechanical oscillation  \cite{moller2017quantum}. The same toolbox enabled entanglement \cite{Thomas2021} and strong coherent coupling \cite{karg2020light} between a membrane and an atomic ensemble. The same type of remote link was employed for quantum coherent measurement and feedback, where reducing entropy removal from the spin degrees of freedom in an atomic ensemble by optical pumping can be converted to cooling of a mechanical mode of vibration
\cite{Schmid2022}.

Common to those setups is that they require interferometric stability to perform quadrature measurements or interact with a phase referenced signal beam. For the interaction with spin systems, this can be quite naturally achieved by encoding quantum information in the polarisation state of a light beam. While one polarisation component serves as a signal, the orthogonal component provides a co-propagating local oscillator. The common path makes this very robust and preserves quantum properties over long distances. However, for the interaction with a mechanical oscillator, the local oscillator field must be separate, requiring either a local \cite{karg2020light} or a distributed interferometer \cite{Thomas2021} and phase stabilisation.
Here we investigate a device that can reduce some of the technical overhead when coupling a polarised light beam to an optical cavity that interacts with a mechanical resonator.

\section{A position to polarization converter}
\begin{figure}[b]
\centering\includegraphics[width=0.9\linewidth]{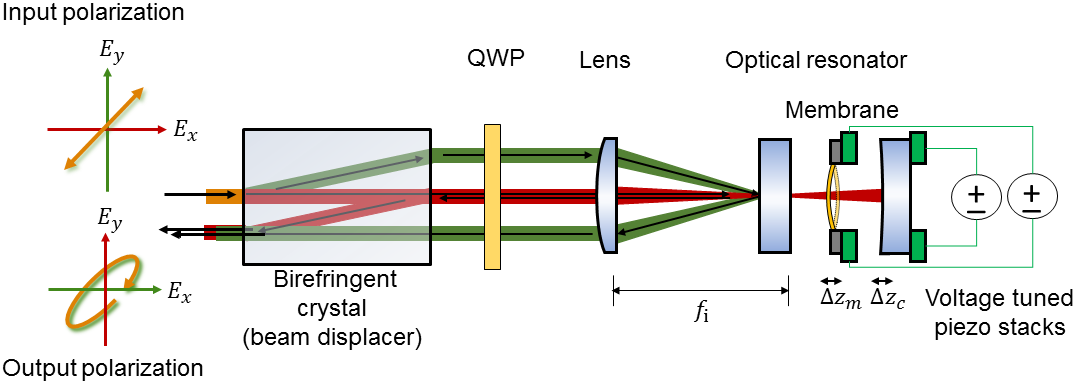}
\caption{Design of the position-to-polarization converter. A stable polarization interferometer is formed between a beam displacer and an asymmetric, plano-convex optical cavity containing a transparent membrane. The input beam is split into signal (red) and reference (green) beams of orthogonal polarization. 
The signal beam is mode-matched to a TEM$_{00}$ mode of the cavity. Upon reflection off the entrance mirror, located in the focal plane of a lens, both beams are recombined by the displacer, thus closing the interferometer. A double-pass through a quarter-wave plate (QWP) provides the necessary exchange of horizontal and vertical polarization components for beam recombination. As a result, membrane motion causes phase shifts of the signal beam and thus variation of the output polarization. Longitudinal displacements of back mirror ($\Delta z_c$) and membrane ($\Delta z_m$) can be actively controlled.}
\label{p-pConverter}
\end{figure}

Our principal design is shown in Fig.~\ref{p-pConverter}. It is based on a polarizing beam displacer/combiner, which spatially splits a polarized input laser beam into two orthogonally polarized components. Both ordinary and extraordinary beams are focused by an imaging lens onto an asymmetric optical cavity. The extraordinary beam is reflected off the first mirror at an angle and serves as an optical phase reference. The ordinary beam travels along the optical axis and is brought into resonance with a cavity mode by tuning the position of the back mirror with a piezo element. The light that enters the cavity interacts with a transparent, micro-mechanical membrane, which causes a significant phase shift that depends on the membrane’s position.
The cavity mirrors have different reflectivity such that the ordinary signal beam leaves the cavity predominantly through the front mirror and can be recombined with the imaged reference beam to form a single output beam. Position changes and oscillations of the membrane are translated into modulation of the resulting polarization, which can be observed by polarimetry or can be used to interact with another polarisation-sensitive system such as dispersively coupled atomic ensembles. Backaction onto the membrane motion arises from changes or fluctuations of the input polarization, which translates into variations of signal beam intensity and radiation pressure inside the cavity.

\section{Beam alignment and mode overlap}
\begin{figure}[b]
\centering\includegraphics[width=0.67\linewidth]{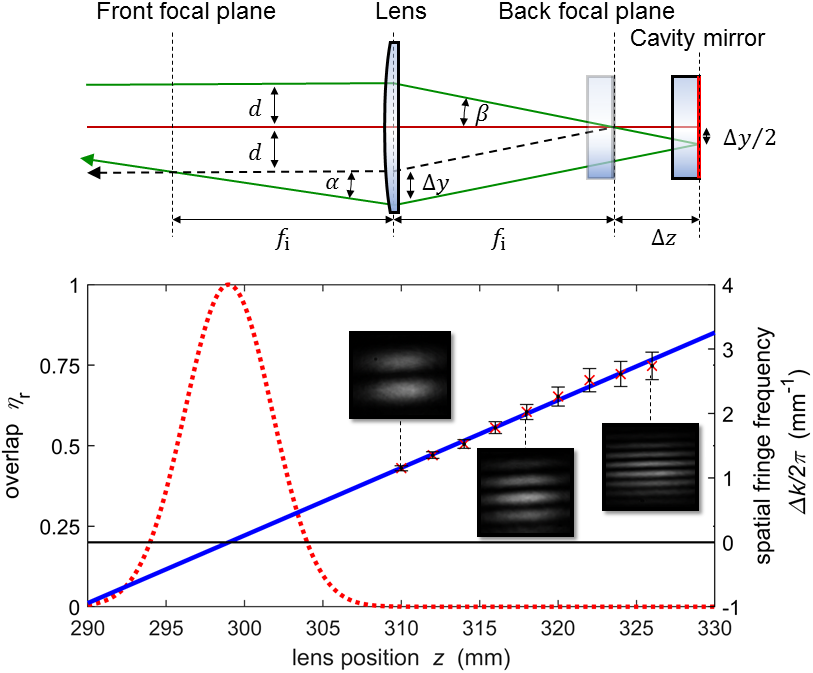}
\caption{Geometry of reference beam alignment (top). If the lens-cavity distance deviates from one focal length, the returning signal (red, along the optical axis) and reference beams (green) will not be parallel, leading to reduced mode overlap after recombination by the beam displacer (not shown). The dashed line shows the intended reference beam path. The depicted degree of typical misalignment is exaggerated here for clarity. Alignment precision and mode overlap (bottom). The spatial fringe frequency is proportional to the deviation of lens-cavity distance from one focal length and can be used to locate the correct lens position (fitted solid blue line). The expected mode overlap for deviations from that position shows the tolerance to misalignment (red dotted profile). The estimated positioning accuracy of $\approx \pm0.7~\text{mm}$ allows for a mode overlap of better than 0.985.
\label{interference}
}
\label{alignment}
\end{figure}
In order to achieve good opto-mechanical coupling between membrane and beam polarization, the input beam must be mode matched to a $\text{TEM}_{00}$ mode of the optical resonator while at the same time ensuring mode-overlap between signal and reference beam paths. 
Mode-matching to the cavity requires control of beam size and divergence, but these should be adjusted by shaping the laser beam before entering the beam displacer.
The distance between the imaging lens and the reflecting surface of the front mirror of the cavity must be equal the focal length $f_\text{i}$ to generate parallel and correctly spaced signal and reference beams returning to the beam displacer. For a collimated beam of the correct diameter and a plano-convex cavity as used here, this condition may coincide with matching the beam divergence to the cavity mode (for vanishing Gaussian focal shift \cite{Sucha1984}).

We can assess the geometric precision that is required for sufficient beam overlap, i.e.\ the tolerance to deviations of the correct lens-mirror distance.
For a distance error $\Delta z$ between the reflecting surface and the back focal plane of the lens, the reflected reference beam will be parallel displaced by a distance $\Delta y=2\Delta z\tan\beta$ from its intended path as illustrated in Fig.~\ref{alignment}. It is straightforward to see that the focusing angle $\beta$ is determined by beam separation $d$ and focal length $f_\text{i}$ according to $\tan\beta=d/f_\text{i}$, and therefore $\Delta y=2\Delta z d/f_\text{i}$. After re-collimation by the lens, the returning reference beam will deviate by a small angle $\alpha$ and cross the intended beam path in the front focal plane. Consequently, we find an angular deviation according to $\tan\alpha=\Delta y/f_\text{i}=2\Delta z d/f_\text{i}^2$.

When signal and reference beam are recombined, a reduced mode overlap manifests as a spatial modulation of resulting beam polarisation in the near field and may lead to beam separation in the far field. For small angles $\alpha$, the beam displacer introduces identical displacements $d$ during the splitting and recombination processes. Therefore, we can evaluate the transversal mode matching by the overlap between the actual and intended reference beam. In the front focal plane of the imaging lens, the two returning beams intersect and differ only in their transverse momentum, given by the wave number difference
\begin{equation}
\Delta k=\frac{2\pi}{\lambda}\sin \alpha\approx2\pi\frac{2 d}{\lambda f_\text{i}^2}\Delta z, 
\label{interference8} 
\end{equation}

Observed interference fringes are shown for different values of $\Delta z$ in Fig.~\ref{interference}.
The spatial fringe frequency increases proportional to distance as the lens is displaced from the focal distance. It vanishes in the vicinity of the nominal focal length of $300~\text{mm}$. From the extrapolation, we can determine the ideal lens position with an uncertainty of $\Delta f_\text{i}\approx \pm0.7~\text{mm}$.

To calculate the mismatch in the overlap between the signal and reference beam, we assume two near-collimated Gaussian beams of a sufficient diameter such that the beams' radii of curvature approach infinity and local changes in wavelength due to the shifting Guoy phases can be neglected. We also assume a constant beam waist $w$ along the collimated beam path, such that the overlap integral between the ideal and tilted reference beam becomes
\begin{equation}
    \eta_\text{r}=\int_{-\infty}^\infty\int_{-\infty}^\infty \frac{2}{\pi w^2}e^{-\frac{x^2+y^2}{w^2}}e^{-i \Delta k x}dx dy=e^{-\frac{\Delta k^2 w^2}{8}},
\label{overlap}
\end{equation}
where $2w$ is the $1/e^2$-width of the beams. For a beam that is mode-matched to our cavity with $\lambda=795~\text{nm}$, a beam waist of $w=0.7~\text{mm}$, a beam displacement of $d=4~\text{mm}$, and a focal length of $f_\text{i}=300~\text{mm}$, we find that an accuracy of $\approx\pm 1.3~\text{mm}$ for positioning the lens is sufficient to achieve a mode overlap between signal and reference beams of better than 95$\%$. 

Specific to our setup, we also estimate the signal beam overlap with the TEM$_{00}$ mode of the present cavity by imaging the reflected signal beam for both resonant and off-resonant conditions. Fitting two-dimensional Gaussian beam model functions including beam tilt and relative divergence to match the two profiles results in a mode overlap of $\eta_\text{c}\approx 0.93$.

\begin{figure}[!ht]
\centering\includegraphics[width=0.9\linewidth]{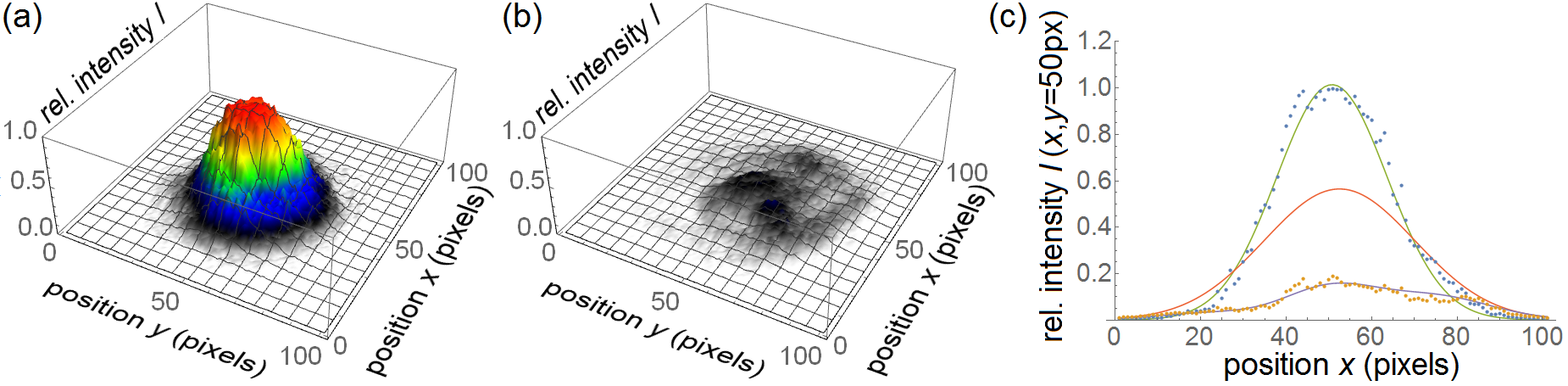}
\caption{Intensity profiles of the reflected signal beam. The near-Gaussian intensity profile of the off-resonant input beam is shown in (a), the residual reflected intensity from the cavity on resonance is shown in (b). Cuts through the data and corresponding modelled profiles as well as the inferred intensity profile of the cavity mode (solid line without data) are shown in (c).}
\label{fig:overlap}
\end{figure}

\section{Polarimetric measurement}
\begin{figure}[b]
\centering\includegraphics[width=0.8\linewidth]{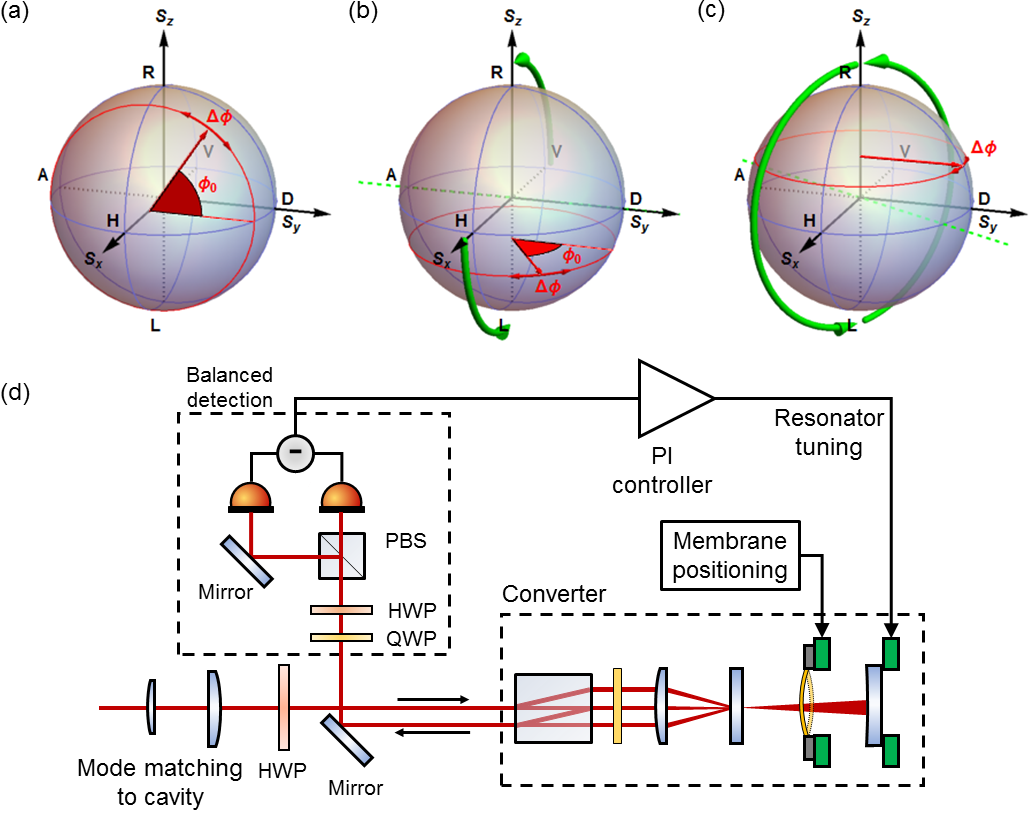}
\caption{
Polarimetric setup and adjustment of measurement basis. The imbalanced signal and reference beam returning from the cavity are combined in a horizontal/vertical (H/V) basis with some phase offset $\phi_0$ and varying phase difference $\Delta\phi$. The combined polarisation is depicted on the Poincar\'e sphere in (a). For varying phase differences, the Stokes vector rotates about the $S_x$-axis. A quarter-wave plate introduces a $90^\circ$ rotation about the anti-/diagonal (A/D) axis, shown in (b), transforming signal and reference to orthogonal circular polarisations (L/R) and thus converting the signal into rotation about the $S_z$-axis. Finally, a half-wave plate allows for the compensation of the phase offset by introducing opposite phase shifts to the circular components (and swapping L/R). As a result, the phase difference $\Delta\phi$ is mapped onto the imbalance between H and V polarisations and measured by the detector pair. An overview of the optical setup is shown in (d).
}
\label{polarimetry}
\end{figure}
The converter acts as a linear-birefringent and linear dichroic reflector. Dichroism arises from loss in the cavity and also depends on the position of the membrane, but we want to predominantly observe birefringence. The beam displacer produces and recombines (unbalanced) linearly polarized signal and reference beams, which we denote as vertical (V) and horizontal (H) components. One of these enters the resonator, and their relative phase carries the variable signal $\Delta\phi$. The phase may include an offset $\phi_0$, arising e.g.\ from imperfect lens alignment or any spurious birefringence. We use homodyne detection, which is now available in the form of a balanced polarimeter, i.e.\ a photo-detector pair illuminated by the two outputs of a polarizing beam splitter, as shown in Fig.~\ref{polarimetry}. 
We use a combination of quarter-wave and half-wave plates to achieve intensity balance between the two detectors and compensate for any phase offset. This may be understood by depicting the polarization output as a vector on the Poincar\'e-sphere, also depicted in  Fig.~\ref{polarimetry}.  
The two wave plates introduce rotations of the sphere such that in first order the intensity imbalance between resulting H and V components becomes directly and maximally proportional to $\Delta\phi$.

More formally, and ignoring any intermediate forward and backward transformations of Stokes vector coordinates due to a non-zero, but compensated phase offset $\phi_0$, the (quantum) measurement represents a detection of the Stokes vector component $\hat S_z$ (difference between circular polarization components) returning from the converter, while the phase shift in the converter introduces rotations about the $S_x$-axis (difference between horizontal and vertical components). Loss of power in the signal beam due to the linear dichroism results in changes of the (unobserved) $\hat S_x$-component and overall power. 

\begin{figure}[t]
\centering\includegraphics[width=0.6\linewidth]{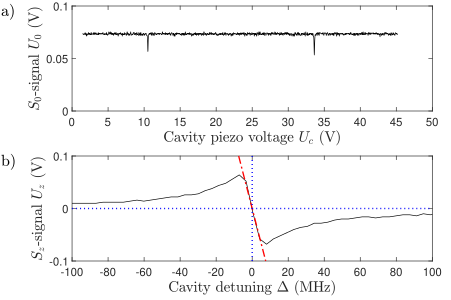}
\caption{Typical polarimeter signals for a scan across optical resonance. The total reflected light power arriving at the detector (sum of reference and signal beams) shows increased absorption at $\text{TEM}_{00}$ cavity resonances (top). The free spectral range can be used to calibrate detuning between cavity and laser frequency. 
The dispersive signal in the vicinity of a resonance can be used for active stabilisation of cavity resonance (bottom). A linear fit to the central slope can be used for sensitivity calibration.}
\label{dispersive}
 \end{figure}

Due to the geometric robustness and nearly identical path lengths in the interferometer (white light condition for off-resonant light, achieved at correct lens position), the polarization output of the converter is sufficiently stable to use the resulting signal from the balanced photo-detector directly for active stabilisation of the optical resonator. This is demonstrated in Fig.~\ref{dispersive}, which shows typical signal responses to varying the cavity length across an optical resonance. The dispersive signal can be fed as an error signal to a proportional/integral (PI) controller to maintain optical resonance. It can also be used to calibrate the polarimeter's signal response to small deviations from resonance, such as caused by thermal and quantum noise of membrane oscillations. See below for an analysis of the corresponding signals.

\section{Membrane in an asymmetric cavity}
The membrane-cavity system that we use here is asymmetric in several senses. We use a plano-convex cavity, the membrane may not be placed near the centre, and the reflectivity of front and back mirrors differs.
We use a simplified, one-dimensional model to describe the behaviour of the system, similar to treatments for specific cases described with explicit expressions elsewhere, e.g.\ for a fully symmetric, membrane-in-the-middle (MIM) scenario \cite{thompson2008strong,jayich2008dispersive}
or membrane-at-the-edge (MATE) cases \cite{dumont2019flexure}
and their behaviour near degenerate points with different mirror reflectivity \cite{dumont2022asymmetry}.

Each optical element can be described by a beam splitter matrix \cite{agarwal2012quantum}. For most purposes, we can ignore specific reflection and transmission phases, which are equivalent to small changes in path lengths between elements. Therefore, we use a symmetric matrix to link the forward and backward travelling complex field amplitudes before ($F_n$ and $B_n$) and after ($F_n^\prime$ and $B_n^\prime$) the $n^\text{th}$ element according to
\begin{align}
\arraycolsep=2pt\def\arraystretch{1.2}
    \left(\begin{array}{c}
         F_n^\prime\\
         B_n
    \end{array}\right)
    =\left(\begin{array}{cc}
       t_n & r_n\\
        r_n & t_n
    \end{array}\right)
    \left(\begin{array}{c}
         F_n\\
         B_n^\prime
    \end{array}\right)
    =\left(\begin{array}{cc}
       i\cos\theta_n & \sin\theta_n\\
        \sin\theta_n & i\cos\theta_n
    \end{array}\right)
    \left(\begin{array}{c}
         F_n\\
         B_n^\prime
    \end{array}\right),
    \label{eq:BSmatrix}
\end{align}
where $\theta_n$ determines lossless transmittance $T_n=\cos^2\theta_n$ and reflectance $R_n=\sin^2\theta_n$. In principle, the matrix elements can also include losses upon transmission and reflection from an element \cite{barnett1998quantum}.
To approach this implicit equation, the effect of any chain of subsequent elements on forward and backward travelling amplitudes can be summarized by 
\begin{equation}
    B_n^\prime=\rho_n^\prime F_n^\prime=\rho_{n+1}\gamma_n e^{2ikL_n} F_n^\prime,
\end{equation}
where $\rho_{n+1}$ is an effective, complex reflection coefficient of the subsequent chain. We distinguish $\rho_n^\prime$ by including the round trip propagation phase to the next element over distance $L_n$ for wave number $k=2\pi/\lambda$ as well as corresponding propagation losses using $\gamma_n$. Combination with Eq.~(\ref{eq:BSmatrix}) then leads to effective transmission and reflection coefficients for the $n^\text{th}$ element according to
\begin{align}
  \tau_n=\frac{F_n^\prime}{F_n}=\frac{t_n}{1-r_n\rho_n^\prime }
  \qquad\text{and}\qquad
  \rho_n=\frac{B_n}{F_n}=\frac{r_n-\rho_n^\prime}{1-r_n\rho_n^\prime }.
\end{align}
These expressions can therefore be chained, using a zero reflection coefficient for the empty space after the final element. The total reflection coefficient of the system is just the effective reflection coefficient $\rho_1$ of the first element, while the total transmission is given by the product of all effective transmission coefficients $\tau_n$, the total propagation phase and all single-pass loss factors. 

In the present setup, we use an optical cavity in a thermally compensated mount (combination of invar and aluminium)  placed under high vacuum. The total cavity length $L+\Delta z_c\approx 3~\text{cm}$ corresponds to free spectral range of $\nu_\text{fsr}\approx 5~\text{GHz}$. The mirror reflectances of $R_1=0.99$ and $R_3=0.9995$ lead to a theoretical finesse for the empty resonator of $\mathcal{F}=595$.
The membrane divides the resonator into two coupled sub cavities of lengths $L_1-\Delta z_m$ and $L_2+\Delta z_m+\Delta z_c$.  For our $\text{Si}_3\text{N}_4$ membrane of $d=50~\text{nm}$ thickness and refractive index $n\approx2.0245$ at $\lambda=795~\text{nm}$ \cite{luke2015broadband}, we estimate reflectance and transmittance with minor absorption losses of \cite{jayich2008dispersive}
\begin{align}
    R_2&\approx\frac{(n^2-1)^2}{(n^2+1)^2+4 n^2 \cot^2(n k d)}\approx0.232.\\
   T_2&\approx 1-R_2-1.6\times 10^{-4}
\end{align}
For the current system with a planar front mirror, we disregard absorption and scattering losses but consider an efficiency $\gamma_1<1$ due to the dominating optical instability of the plane-parallel sub-cavity, which depends strongly on the parallel alignment of the membrane and front mirror. We assume $\gamma_2=1$, i.e.\ no further losses in the back cavity other than the transmission through the back mirror. 

\begin{figure}[htbp]
\centering\includegraphics[width=0.75\linewidth]{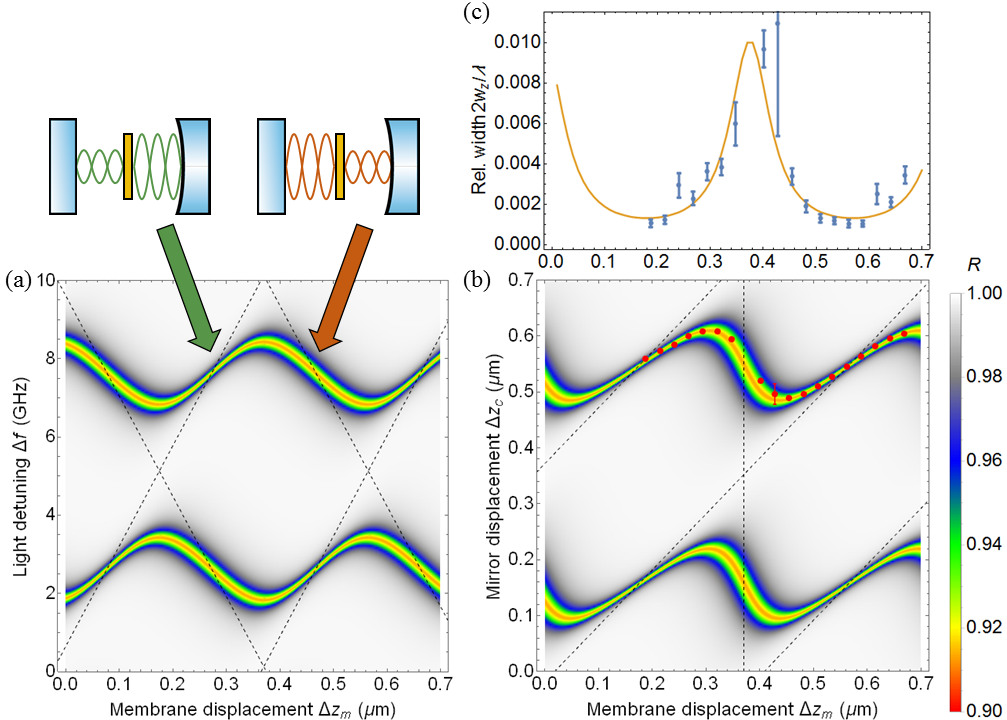}
\caption{Simulations show the total reflectance of the resonator as a function of the membrane position and laser frequency (a) or back mirror position (b). For clarity, an exaggerated optical instability loss with $\gamma_1=0.8$ was used, resulting in low-finesse modes. Dashed lines indicate lossless sub-cavity modes for 100\% membrane reflectivity. In (b), experimental data for resonance locations has been overlayed (red dots). Corresponding full linewidths, measured as mirror displacement $w_z$, are shown in (c) together with the theoretical expectation for an estimated $\gamma_1\approx0.994$. We attribute deviations to spurious coupling to higher-order, transversal cavity modes.}
\label{fig:tuning}
\end{figure}

Fig.~\ref{fig:tuning} shows the theoretical cavity response according to this model as a function of membrane position $\Delta z_m$ for tuned laser frequency as well as for tuned back mirror position $\Delta z_c$. The former might be intuitively understood in terms of decreasing and increasing mode energies of the two sub-cavities with different finesse. 
Their coupling by transmission through the membrane leads to avoided crossings. As a result, resonant frequencies as well as cavity finesse
oscillate as a function of membrane displacement $\Delta z_m$. This behaviour corresponds to the maximum field intensity alternating between both sub-cavities.
For the case of tuned mirror position, one should note that the mirror position does not influence the length of the front sub-cavity and is thus not equivalent to tuning the laser frequency. Here, the apparent linewidth of a resonance is determined by a mixture of cavity decay rate and tuning behaviour. 
This case is used for comparison with experimental data also shown in Fig.~\ref{fig:tuning}, where we used a fixed laser frequency referenced to an atomic transition within the $^{87}$rubidium $D_1$-line manifold. 

The strongest dispersive coupling to the membrane occurs when the back sub-cavity is resonant, while the front sub-cavity is anti-resonant ($L_1=\lambda/4+m \lambda/2$ and $L_2+\Delta z_c=n \lambda/2$ with integer $m,n$ for the same sign of all $r_i$), see indicating arrow in Fig.~\ref{fig:tuning} (a). Here, the first-order expansion of the effective reflection coefficient becomes
\begin{align}
\rho_1\approx\frac{r_1(r_2r_3-1)+\gamma_1(r_3-r_2)}{(r_2r_3-1)+r_1\gamma_1(r_3-r_2)}+4\pi i \frac{(r_1^2-1)r_2\gamma_1(2r_2r_3-1-r_3^2)}{((r_2r_3-1)+r_1\gamma_1(r_3-r_2))^2}\frac{\Delta z_m}{\lambda},
\label{eq:response}
\end{align}
where all amplitude reflection coefficients $r_{1,2,3}$ and $\gamma_1$ are real.
One should note here that the reflected intensity will reach a minimum on resonance and may completely vanish, i.e.\ lead to $\rho_1(\Delta z_m=0)=0$. This impedance-matching condition is given by
\begin{align}
    r_1=\frac{\gamma_1(r_2-r_3)}{r_2 r_3-1},
\end{align}
which reduces to $r_1\approx\gamma_1$ for highly reflective back mirrors with $r_3\approx 1$. However, the signal response is entirely given by the imaginary part of the expression in Eq.~(\ref{eq:response}), which also corresponds fully to the light quadrature that is measured by the correctly balanced homodyne detector.
For fixed loss, this response reaches its maximum exactly for the same condition. If maximal response is required, the front mirror reflectivity should thus be chosen for impedance matching \cite{Chow2008}.
The response then increases monotonously with $\gamma_1$ and reduces to 
\begin{align}
    \rho_1\approx i\chi\Delta z_m= 8\pi i\frac{\gamma_1}{1-\gamma_1^2}\frac{r_2}{1-r_2}\frac{\Delta z_m}{\lambda}
\end{align}
for $r_3\approx 1$. This expression diverges for $\gamma_1\rightarrow 1$ or $r_2\rightarrow 1$, which would both correspond to infinitely sharp resonance width. In scenarios where the interaction with the membrane should ideally be lossless, in particular to retain the quantum properties of the input beam, the resonator should be undercoupled with a front mirror reflectivity much lower than the internal loss factor. For zero loss with $\gamma_1=r_3=1$, the response would be given by
\begin{align}
    \rho_1\approx-1+8\pi i\frac{1+r_1}{1-r_1}\frac{r_2}{1-r_2} \frac{\Delta z_m}{\lambda}
\end{align}

Experimentally, we find an optical instability loss with $\gamma_1\approx0.994$ in our present system that leads to operation close to the impedance matching condition $\gamma_1^2\approx R_1$.

\section{Noise measurements}

To assess the suitability of our setup for quantum optical experiments, we observe and evaluate levels of measurement noise. In particular, we observe thermal membrane motion and compare it against laser frequency noise.

The polarimetric detector measures the
Stokes vector component $\hat S_z$. Here, we follow the definitions for photon flux differences in a right-handed coordinate system
\begin{align}
\def\arraystretch{1.2}
    \left(\begin{array}{c}
        \hat S_x\\
        \hat S_y\\
        \hat S_z
    \end{array}\right) &=\frac{c}{2}\left(\begin{array}{c}
        \hat n_\text{H}-\hat n_\text{V}\\
        \hat n_\text{D}-\hat n_\text{A}\\
        \hat n_\text{L}-\hat n_\text{R}
    \end{array}\right)
        =\frac{c}{2}\left(\begin{array}{c}
        \hat a^{\dagger}_\text{H}\hat a_\text{H}- \hat a^{\dagger}_\text{V}\hat a_\text{V}\\
        \hat a^{\dagger}_\text{H}\hat a_\text{V}+ \hat a^{\dagger}_\text{V}\hat a_\text{H}\\
        i\hat a^{\dagger}_\text{V}\hat a_\text{H}- i\hat a^{\dagger}_\text{H}\hat a_\text{V}
    \end{array}\right),
\end{align}
where $c$ is the speed of light. The annihilation operators $\hat a_{\text{L},\text{R}}=(\hat a_\text{H}\mp i\hat a_\text{V})/\sqrt{2}$, $\hat a_{\text{D},\text{A}}=(\pm\hat a_\text{H}+\hat a_\text{V})/\sqrt{2}$, and $\hat a_\text{H,V}$ describe left/right-handed circular, 
linear diagonal/anti-diagonal, and linear horizontal/vertical beam polarisations, respectively. These Heisenberg operators for field amplitude obey $[\hat a_i(z),\hat a^\dagger_j(z^\prime)]=\delta_{i,j}\delta_z(z-z^\prime)$ for orthogonal polarizations $i,j$, such that the number operators $\hat n_i=\hat a^\dagger_i \hat a_i$ describe linear spatial photon density. Observing that $\delta(t)=c\delta_z(z=ct)$, the Stokes vector obeys angular momentum commutation rules according to $[\hat S_x(t),\hat S_y(t^\prime)]=i\delta(t-t^\prime)\hat S_z(t)$ and cyclic permutations. The total photon flux is given by
$\hat{\Phi}=2\hat S_{0}=c(\hat n_\text{H}+\hat n_\text{V})= c(\hat a^\dagger_\text{H}\hat a_\text{H}+\hat a^\dagger_\text{V}\hat a_\text{V})$.   

We illuminate the converter with a coherent, linearly polarized input beam under a an angle $\alpha$, such that the input polarization is described by $\langle\hat{a}_\text{H,in}\rangle=A\cos{\alpha}$,  $\langle\hat{a}_\text{V,in}\rangle=A\sin{\alpha}$, and thus $\langle\hat{S}_{z,\text{in}}\rangle=0$, $\langle\hat{S}_{y,\text{in}}\rangle=\langle\hat{S}_{0,\text{in}}\rangle\sin{2\alpha}$, and an imbalance between reference (H) and signal (V) beams given by $\langle\hat{S}_{x,\text{in}}\rangle=\langle\hat{S}_{0,\text{in}}\rangle(\cos^2\alpha-\sin^2\alpha)$. The effect of the converter is that of an effective beam splitter acting on the partially entering signal field, while the reference beam is fully reflected. Ignoring the exchange of horizontal and vertical polarisations, the detected field is thus given by
\begin{align}
    \hat{a}_\text{H}=\hat{a}_\text{H,in}\qquad\text{and}\qquad\hat{a}_\text{V}=\rho_1\hat{a}_\text{V,in}+i\sqrt{1-|\rho_1|^2}\hat{a}_\text{V,0},
\end{align}
where $\hat{a}_\text{V,0}$ is a vacuum field that arises from the losses in the cavity. We assume impedance matching and resonance with the cavity such that $\rho_1\approx i\chi\Delta z_m$. We also assume $\langle\Delta z_m\rangle=0$ and the coupling to membrane motion to be small enough such that $|\rho_1|^2\ll1$.
As a consequence, the measurement is described by
\begin{align}
    \hat{S}_z\approx\frac{c}{2}\left(\chi\Delta z_m(\hat{a}^\dagger_\text{V,in}\hat a_\text{H,in}+ \hat a^{\dagger}_\text{H,in}\hat{a}_\text{V,in})+(\hat{a}^\dagger_\text{V,0}\hat a_\text{H,in}+ \hat a^{\dagger}_\text{H,in}\hat{a}_\text{V,0})\right)
\end{align}
Writing the field operators at
the input as $\hat{a}_\text{H,V}=\langle \hat{a}_\text{H,V}\rangle+\delta \hat{a}_\text{H,V}$ and neglecting all higher-order terms results in
\begin{align}
    \hat{S}_z(t)\approx\eta\langle\hat{S}_{y,\text{in}}\rangle\chi\Delta z_m(t)+\sqrt{\eta\frac{c}{2}\langle\hat{S}_{0,\text{in}}\rangle}\cos\alpha(\hat{a}^\dagger_\text{V,0}(ct)+\hat{a}_\text{V,0}(ct)),
\end{align}
where we included a detection loss $\eta$, which reduces the photon flux but replenishes the vacuum field.
The first term measures the membrane displacement while the second term arises from the co-measured field quadrature of the vertically polarised vacuum field, which leads to white photon shot noise. This signal is measured with a  frequency-dependent electronic gain $g_\text{el}$ such that $U(t)=g_\text{el} S_z(t)$.
 Since membrane position and vacuum field are uncorrelated, the auto-correlation function $R_{UU}(\tau)=g_\text{el}^2\langle\hat{S}_z(t)\hat{S}_z(t+\tau)\rangle$ of the measured voltage is given by
\begin{align}
    R_{UU}(\tau)  =g_\text{el}^2\eta^2\langle\hat{S}_{y,\text{in}}\rangle^2\chi^2R_{zz}(\tau)+\frac{1}{2}g_\text{el}^2\eta\cos^2\alpha\langle\hat{S}_{0,\text{in}}\rangle \delta(\tau),
\end{align}
where we introduced the auto-correlation  $R_{zz}(\tau)=\langle\Delta z_m(t)\Delta z_m(t+\tau)\rangle$ of membrane motion.
Without electronic noise, the power spectral density of the measured voltage is given by
\begin{align}
    S_{UU}(f)=\int_{-\infty}^{\infty}R_{UU}(\tau)e^{-2\pi i f \tau}d\tau=g_\text{el}^2\eta^2\langle\hat{S}_{y,\text{in}}\rangle^2\chi^2 S_{zz}(f)+\frac{1}{2}g_\text{el}^2\eta\cos^2\alpha\langle\hat{S}_{0,\text{in}}\rangle.
    \label{eq:noise}
\end{align}
Here, the power spectral density (per natural frequency) of an underdamped,  thermally driven oscillator with resonant frequency $f_0$ and energy loss rate $\gamma$ is approximately given by
\begin{align}
    S_{zz}(f)=\int_{-\infty}^\infty R_{zz}(\tau)e^{-2\pi i f \tau}d\tau\approx\frac{k_BT}{m_\text{eff}(2\pi f_0)^2} \frac{2\gamma}{16\pi^2 (|f|-f_0)^2+\gamma^2},
\end{align}
where $m_\text{eff}$ is the effective mass taking part in the oscillation.

\begin{figure}[tb]
\centering\includegraphics[width=0.8\linewidth]{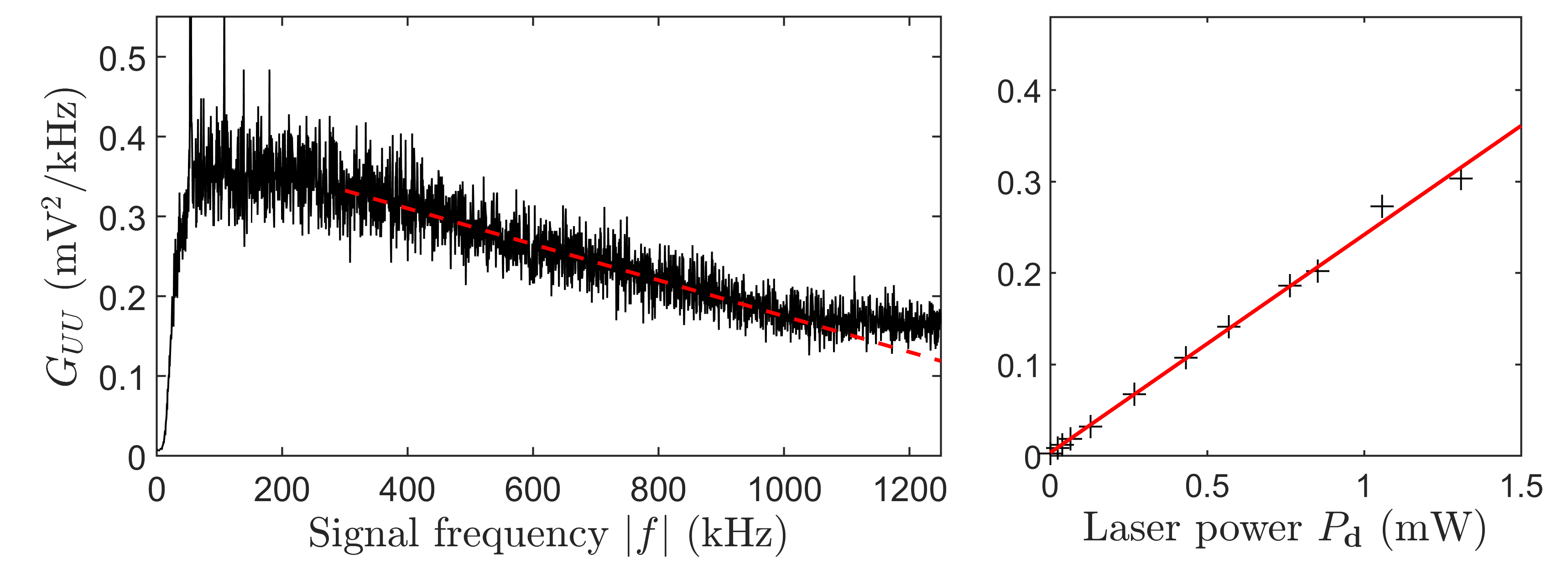}
\caption{Shot noise limited detector response. The spectral response of our $1~\text{MHz}$-bandwidth balanced detector to photon shot noise is shown in (a) for a power arriving at the detector of $P=1.3~\text{mW}$. One-sided power spectral density $G_{UU}(|f|)=2S_{UU}(f)$ was measured with $500~\text{Hz}$ effective bandwidth (100 averages of 2~ms data). The response below $\approx 80~\text{kHz}$ is limited by an additional high-pass filter. We typically observe spurious narrowband signals in the region below $100~\text{kHz}$. Above $1~\text{MHz}$, excess noise due to aliasing of higher frequencies at $2.5~\text{MHz}$ sampling rate becomes visible. A linear extrapolation of the falling slope (red dashed line) is shown as an indication. The linear scaling with laser power (photon flux) is shown in (b) for a signal frequency of $400~\text{kHz}$. We find a slope of $G_{UU}/P_\text{d}=2.4\times 10^{-7}~\text{V}^2\text{Hz}^{-1}\text{W}^{-1}$ with an offset of $4.6\times 10^{-12}~\text{V}^2\text{Hz}^{-1}$ due to electronic noise. The slope corresponds to an electronic gain of $g_\text{el}=\sqrt{4eG_{UU}/ rP_\text{d}}\approx~ 5.24\times10^{-13}~\text{V/Hz}$, assuming a detector responsivity of $r=I/P_\text{d}=0.56~\text{A/W}$ at $\lambda=795~\text{nm}$ corresponding to a quantum efficiency of $\eta=r hc/e\lambda=0.88$.}
\label{fig:shotnoise}
\end{figure}

For signal detection, we use a balanced detector pair (Thorlabs model PDB210A) with shot noise-limited performance over its $\approx 1~\text{MHz}$ bandwidth, see Fig.~\ref{fig:shotnoise}. The shot noise level allows for calibration of the electronic gain $g_\text{el}$, using Eq.~(\ref{eq:noise}) and the photon flux arriving at the detector $\Phi_\text{d}=2\cos^2\alpha\langle\hat{S}_0\rangle=P_\text{d}\lambda/hc$ measured as light power $P_\text{d}$.

\begin{figure}[tb]
\centering\includegraphics[width=0.9\linewidth]{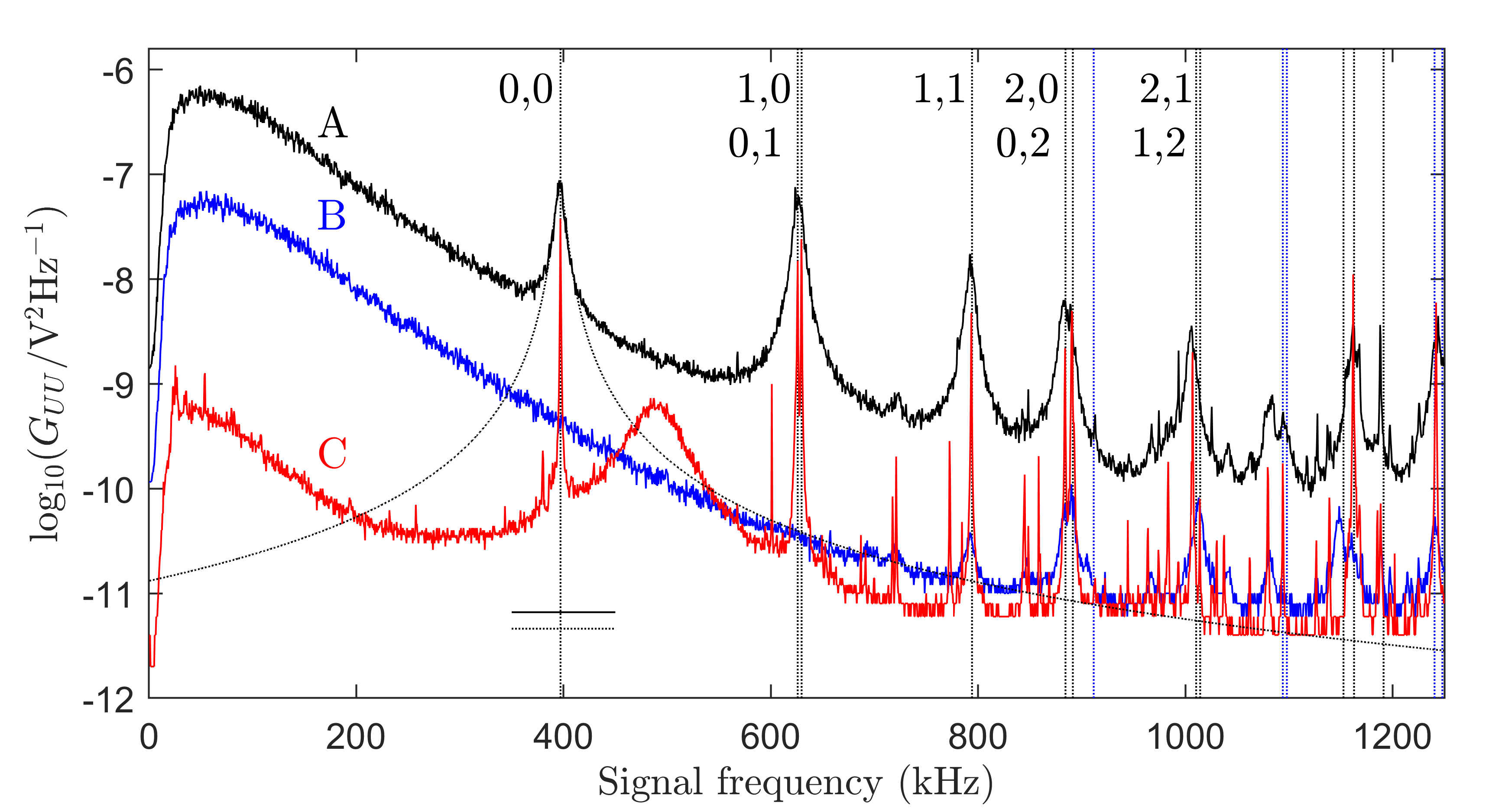}
\caption{Thermal noise of the membrane is detected for dispersive coupling (trace A). A model fit to the fundamental membrane mode is indicated with a profile. For membrane positions near minimal dispersive coupling (trace B) only higher order membrane modes remain visible. Here, the combined level (black line) of photon shot noise ($\approx2.0\times10^{-12}~\text{V}^2/\text{Hz}$) and electronic noise ($\approx4.6\times10^{-12}~\text{V}^2/\text{Hz}$) near $400~\text{kHz}$ at a total detected light power of $8.3~\mu\text{W}$ is just above the electronic noise (dashed line). For improved conditions (trace C) with lower membrane damping (lower vacuum pressure) and reduced laser frequency noise, signal-to-noise-ratio improves and near-degenerate modes can be resolved. This trace was obtained with lower laser power $P=6.7~\mu\text{W}$ and smaller polarisation angle $\alpha\approx16^{\circ}$.
For comparison, (aliased) expected mode frequencies for an almost square membrane (1.01 side ratio) with a fundamental frequency of $397.1~\text{kHz}$ are indicated.}
\label{fig:dispersivedissipative}
\end{figure}

When the optical cavity is locked on resonance (here done by stabilizing the cavity length), the power spectral density of the polarimeter signal exhibits distinct features that we can identify as thermally excited modes of oscillation of the square membrane (Brownian motion), see Fig.~\ref{fig:dispersivedissipative}. 
The frequencies are consistent with the modes of an almost square membrane, matching the expected fundamental frequency $f_0=\sqrt{2T/\rho}/2a\approx397~\text{kHz}$ of the thin
stoichiometric $\text{Si}_3\text{N}_4$ membrane under a tensile stress of $T\approx1.0~\text{GPa}$ with a density $\rho\approx3.17~\text{g/cm}^3$ and dimensions $a\times a\times b=1~\text{mm}\times1~\text{mm}\times50~\text{nm}$.
Due to relatively high residual vacuum pressure, the damping rate of the membrane oscillations is relatively high in this measurement, with a full width half maximum 
(FWHM) $\gamma\approx2\pi\times10.3~\text{kHz}$.
From a model fit to the experimental data, we find a contribution of the membrane's fundamental mode to the variance of the signal voltage of $\sigma^2=1.25\times10^{-3}~\text{V}^2$. It compares very well with the expected thermal variance of $\sigma_{f_0}^2=g_{\text{el}}^2\eta^2\sin2\alpha\langle\hat{S}_{0,\text{in}}\rangle^2\chi^2 k_B T/m_\text{eff}(2\pi f_0)^2\approx 1.4\times 10^{-3}~\text{V}^2$ for $m_\text{eff}=\rho a^2 b/4$, $T=300K$ an input power of $P\approx 11.1~\mu\text{W}$ with a measured polarisation angle of $\alpha\approx 37.5^{\circ}$. The observed variance is lower than theoretically estimated as we did not include reduced mode overlap and transversal misalignment of the membrane with respect to the optical $\text{TEM}_{00}$ mode.
The membrane features mostly vanish when the membrane is positioned at points of minimal dispersive coupling.
Some higher-order membrane modes remain visible, which is likely due to weakly coupled transversal cavity modes. 

Compared to room-temperature thermal noise, the variance of the membrane's quantum fluctuations will be approximately 6 orders of magnitude smaller. In the present setup, these are masked by coloured broadband noise, well above the photon shot noise level. It results from frequency fluctuations of the illuminating laser. 
An analysis similar to the above shows that the first-order response to laser frequency fluctuations for maximal dispersive coupling, matched impedance and $r_3\approx 1$ is given by
\begin{align}
    i\chi_f=\frac{\partial\rho_1}{\partial \Delta f}\approx-\frac{4\pi i}{c} \frac{(1-r_2)L_1+(1+r_2)L_2}{1-r_2}\frac{\gamma_1}{1-\gamma_1^2},
\end{align}
and will be suppressed for shorter resonator lengths.
The decrease of this noise for minimal dispersive coupling is consistent with the increase in cavity line width and thus reduced frequency response.
For comparison, Fig.~\ref{fig:dispersivedissipative} shows signal trace for reduced frequency noise. Here, the laser was passed through a filter cavity of $\approx 160~\text{kHz}$ linewidth. Excess noise in the region of $\approx 500~\text{kHz}$ arises from the active stabilisation loop (servo bump). In addition, the vacuum pressure was reduced, which decreased membrane damping to $\gamma\approx2\pi\times 480~\text{Hertz}$ and thus contributes a factor of $\approx20$ to the improvement in signal-to-noise ratio (SNR). 

This measurement is still in the weak coupling regime (quantum noise signal of the membrane lower than photon shot noise), but since signal power scales quadratically with laser power while photon shot noise increases linearly, the opto-mechanical coupling strength increases with power. Technical frequency noise power also scales quadratically, such that it will be possible to observe quantum noise more easily for a higher-order mode of the membrane's vibrations at higher signal frequencies where the technical noise will drop below the shot-noise level. However, increased light power also leads to backaction noise and cooling or heating of the membrane via radiation pressure. For increased power, we observe an onset of higher-order mode oscillations, which should be avoided by controlling the transverse alignment of the membrane with respect to the cavity modes and the use of an additional cooling laser.

\section{Conclusion}
We demonstrated an interferometrically stable polarisation interferometer that converts the phase shift from an optical resonator into beam polarisation. It can be directly used  to stabilise the laser-resonator detuning without need for laser modulation. Here, we applied it to detecting microscopic motion of a micro-mechanical membrane. Light-membrane interaction at the quantum level should be feasible for sufficiently reduced laser frequency noise or reduced frequency response from a shorter cavity, making this arrangement a useful tool for hybrid quantum systems. Depending on the intended protocol, the cavity design should consider if maximally coupled read-out of membrane motion or minimal optical loss required.
Our current cavity design uses a plane-parallel entrance mirror, which leads to diffraction loss between that mirror and the membrane and results in impedance matched coupling. To minimize optical loss when operating in the under-coupled regime, the front mirror may as well be replaced by a concave mirror to make the sub-cavity optically stable. For a mode-matched input beam, alignment to the reference beam will remain the same as the mirror will act as a field lens in the focal plane of the imaging lens for beam displacement.

\section{Acknowledgements}
The authors gratefully acknowledge financial support by Jazan University, Al Maarefah Rd, Jazan, Saudi Arabia. We also thank V.~Atkocius for experimental support and proof-reading the manuscript.

\bibliography{coupler}

\end{document}